\documentclass{aa}
\usepackage{latexsym}
\usepackage{graphicx}
\usepackage{natbib}

\def\aap{A\&A}
\def\aaps{A\&AS}
\def\apj{ApJ}
\def\apjs{ApJS}
\def\apss{Astrophys. Space Sci.}

\def\jgr{J.\ Geophys.\ Res.}
\def\mnras{MNRAS}
\def\nat{Nat}
\def\solphys{Sol.\ Phys.}
\def\philtrans{Phil.\ Trans.\ Roy.\ Soc.\ Lond.}

\def\araa{Ann.\ Rev.\ Astron.\ Astrophys.}
\newcounter{IonCS}
\renewcommand{\ion}[2]{\setcounter{IonCS}{#2}#1\,{\sc{\roman{IonCS}}}}

\newcommand{\fig}[1]{Fig.\,\ref{#1}}
\newcommand{\tab}[1]{Table\,\ref{#1}}
\newcommand{\sect}[1]{Sect.\,\ref{#1}}

\newcommand{\FWHM}{{\sc{fwhm}}}
\newcommand{\aCenA}{$\alpha$Cen\,A}
\begin{document}



\title{\mbox{First VUV full-Sun spectrum of the transition region}
       \mbox{with high spectral resolution compared to cool stars}} 

\titlerunning{VUV full-Sun spectrum}
\authorrunning{Peter}

\author{Hardi Peter}

\institute{Kiepenheuer-Institut f\"ur Sonnenphysik,
                79104 Freiburg, Germany; {\tt peter@kis.uni-freiburg.de}}

\date{Received 11. October 2005 / Accepted  ...}

\abstract{
%
This paper reports the first full-Sun vacuum ultraviolet (VUV) emission
line profile originating from the transition region from the chromosphere
to the corona.
It is based on a raster scan of the whole solar disk using SUMER/SOHO.
The full-Sun spectrum has a spectral resolution which allows an
investigation of details in the line profile as well as a thorough
comparison to stellar spectra as obtained, e.g.\ with FUSE or STIS/HST.
The full-Sun spectrum shows enhanced emission in the wings, and is well
described by a double Gaussian fit with a narrow and a broad component.
It is shown that the broad component is due to structures on the solar
surface, especially those related to the magnetic chromospheric network.
Thus it is proposed that the broad components of other solar-like stars
are also a consequence of the mixture of surface structures, and
\emph{not} necessarily a signature of small-scale heating processes like
explosive events, as it is commonly argued.
A comparison to spectra of luminous cool stars shows that the line
asymmetries of these stars might also be a surface structure effect and
not or only partly due to opacity effects in their cool dense winds.
These comparisons show the potential of high quality full-Sun VUV spectra
and their value for the study of solar-stellar connections.
As an example, this study proposes that {\aCenA} has a considerably higher
amount of magnetic flux concentrated in the chromospheric magnetic network
than the Sun.
}

\maketitle

\section{Introduction}

Our information on processes in the outer atmospheres of the Sun,
solar-like stars and stars of other types is largely based on the
analysis of emission line spectra from the ultraviolet and extreme vacuum
ultraviolet (VUV) to the X-rays.
Of special value is the VUV wavelength range from the Lyman edge at
912\,{\AA} to 1600\,{\AA}, which is dominated by emission lines formed at
temperatures from $10^4$\,K to almost $10^7$\,K and chromospheric continua.
Through this temperature coverage all the way from the chromosphere into
the corona, the VUV is best suited to investigate the heating processes
and dynamics of coronae and winds of cool stars.

On the instrumental side, in the VUV one can resolve even details of the
emission line profiles.
The typical thermal line width at $10^6$\,K is of the order of
$w_{\rm{th}}{\approx}30$\,km/s, which is setting a minimum requirement for
the spectral resolution of
$\lambda/\Delta\lambda=c_0/w_{\rm{th}}{\approx}10\,000$, with $c_0$ being the
speed of light in vacuum.
This is fulfilled by current VUV spectrographs, but so far X-ray
instruments operating below 100\,{\AA} lack this capability (cf.\
\tab{T:instr}a).

To interpret the stellar observations and to relate features in the
spectral profile to actual physical processes or structures on the stellar
surface, a comparison to observations of the Sun with high spectral,
spatial, and temporal resolution is of vital importance.
To compare results from solar studies with stellar observations, one needs
high spectral resolution spectra of the Sun observed as a star.
However, such a reference VUV spectrum of the Sun is lacking.
Therefore \cite{Pagano+al:2004} went as far as proposing to use an {\aCenA}
spectrum as a reference full-Sun spectrum.
This is based on the assumption that despite its differences to the Sun,
{\aCenA} is solar-like enough to serve this purpose.

There are plenty of solar spectra with sufficient spectral resolution
(cf.\ \tab{T:instr}b), e.g.\ with the SUMER spectrograph on-board SOHO
\citep{Wilhelm+al:1995,Wilhelm+al:1997}, even for different solar
structures  \citep{Curdt+al:2001:atlas,Curdt+al:2004:atlas}.
The largest slit of SUMER covers
$4{\arcsec}\!{\times}300{\arcsec}$.
This corresponds to 0.04\,\% of the visible solar disk, with the slit length
covering about 15\,\% of the disk diameter.
On the other hand, considering typical line widths in the VUV of the
order of 0.1--0.2\,{\AA}, instruments observing the Sun as a star lack the
spectral resolution necessary to resolve the line profile
(cf. \tab{T:instr}c).
Thus there is a real need for a Sun-as-a-star or full-Sun VUV spectrum.

\begin{table}
\caption{%
Selection of present and past VUV spectrographs along with the wavelength
range covered and approximate (mean) spectral resolution (element) given in
absolute units as well as $\lambda/\Delta\lambda$.
The first block (a) gives instruments for stellar studies, the second one
(b) slit instruments for solar observations covering only a small part of
the solar disk with a single exposure, and the final block (c) shows
instruments observing the Sun as a star.
\label{T:instr}}
%
%
\begin{center}
\begin{tabular}{@{}llr@{}c@{\,}ll@{~}rccc@{}}
\hline
\hline
\\[-1.9ex]
&
& \multicolumn{3}{c@{}}{wavelength}
& \multicolumn{2}{c@{}}{\underline{spec.\ resolution}}\\
& instrument
& \multicolumn{3}{c@{}}{range ~ [\AA]}
& [\AA]	& $\lambda/\Delta\lambda$
& ref.
\\[0.5ex]
\hline
\\[-1.5ex]
(a) &	XMM/Newton	&    6&--&38	& 0.05	&     500	& (1) \\
    &	Chandra/LETG	&    5&--&175	& 0.06	&    1000	& (2) \\
    &	FUSE		&  912&--&1180	& 0.05	&  20\,000	& (3) \\
    &	STIS/HST	& 1130&--&3100	& 0.02	& 100\,000	& (4) \\
    &	GHRS/HST	& 1100&--&3200	& 0.02	& 100\,000	& (5) 
\\[0.5ex]
\hline
\\[-1.5ex]
(b) &	CDS/SOHO	&  150&--& 780 	& 0.2	&   3\,000	& (6) \\
    &	SUMER/SOHO	&  465&--&1610	& 0.04	&  20\,000	& (7) \\
    &	HRTS    	& 1190&--&1730	& 0.05	&  30\,000	& (8) \\
    &	UVSP/SMM	& 1150&--&3600	& 0.02	& 100\,000	& (9)
\\[0.5ex]
\hline
\\[-1.5ex]
(c) &	EGS rocket	&  300&--&1100	& 2  	&      400 	& (10) \\
    &	SOLSTICE	& 1190&--&4200	& 1.5	&   2\,000	& (11)
\\[0.5ex]
\hline
\end{tabular}
\end{center}
\vspace{-2ex}
{\footnotesize
References:
 1\,--\,\cite{denHerder+al:2001};~
 2\,--\,\cite{Brinkman+al:2000};~
 3\,--\,\cite{Moss+al:2000};~
 4\,--\,\cite{Woodgate+al:1998};~
 5\,--\,\cite{Brandt+al:1994};~
 6\,--\,\cite{Harrison+al:1995};~
 7\,--\,\cite{Wilhelm+al:1997};~
 8\,--\,\cite{Brueckner+Bartoe:1983};~
 9\,--\,\cite{Woodgate+al:1980};~
10\,--\,\cite{Woods+Rottman:1990};~
11\,--\,\cite{Rottman+al:1993};~
}
\end{table}

One possibility to get line profiles of the Sun-as-a-star might be to
point a spectrograph above the limb and to analyze the instrumental
stray-light.
The major complication here is that the different parts of the solar disk
contribute differently to the stray-light, and the results depend on the
stray-light model used.
Nevertheless, this has been done using SUMER and results have been reported
by \cite{lemaire+al:2004}.
However, they only analyzed continuum intensities, and not details of line
profiles, yet.
Further work will have to show if this technique can provide high
spatial resolution data with sufficient quality.

It is known since long that for a large variety of stars the emission
lines originating from the transition region from the chromosphere to the
corona, i.e.\ plasma at about $10^5$\,K, show enhanced emission in the
wings \citep{Linsky+Wood:1994,Wood+al:1996,Wood+al:1997}.
Recently this was studied in depth for a larger number of emission lines in
the case of {\aCenA} \citep{Pagano+al:2004}.
These profiles are usually well fitted by double Gaussians, with a narrow
component fitting the line cores and a broad component accounting for the
excess in the wings.

Before they were discovered with stars, such wing enhanced profiles have
been found on the Sun \citep{Kjeldseth-Moe+Nicolas:1977}, and have been
interpreted as being the signature of small-scale reconnection events, so
called explosive events \citep{Dere+Mason:1993}.
Recently an alternative scenario was suggested by \cite{Peter:2001:sec} in
which the broad components might be due to waves propagating up into the
corona.

The rough similarity of the double Gaussian line profiles of solar
spectra, usually acquired near disk center, and the disk-integrated stellar
profiles led a number of authors to attribute the broad components of the
stellar profiles also to small-scale reconnection events
\citep{Wood+al:1997,Pagano+al:2004}.
However, there are subtile differences between the stellar and the
spatially resolved solar spectra, especially concerning the relative
contribution of the broad component and its line width, as will be pointed
out later in this paper.

Also emission lines of luminous cool stars show such double Gaussian
profiles, as has been recently noted by
\cite{Dupree+al:2005}.
They proposed a working hypothesis that the line asymmetries are due to an
opacity effect in the  cool dense wind of these stars.
It remains to be seen, if this sound interpretation is unique, or if the
distribution of surface structures could also lead to the observed spectra.
Here, again, a full-Sun line profile would be of vital importance.

In order not to compare apples with pears, one has to compare
the stellar line profiles to an actual Sun-as-a-star or full-Sun spectrum.
Because of the lack of a direct observation, one had to get a proxy for
such a profile.
For their comparison of {\aCenA} and the Sun \cite{Pagano+al:2004} used
solar spectra from UVSP and SUMER (cf.\ \tab{T:instr}) acquired at disk
center.
To get the solar irradiance they corrected the disk center radiance by
simply multiplying with a factor $\pi R_\odot^2$, where $R_\odot$ is the
solar radius \citep[cf.][]{Wilhelm+al:1998}.
Thus they did \emph{not} account for any center-to-limb variation, which
can be more than a factor of 10 for transition region lines \citep[see
Table 6 of][]{Wilhelm+al:1998}.

In their study of the magnetic activity of ${\tau}$\,Cet and {\aCenA} as
compared to the Sun, \citet{Judge+al:2004} properly accounted for the
center-to-limb variation of the radiance by using observed center-to limb
intensity variations from actual observations with the SUMER instrument
by \cite{Dammasch+al:1999:ctl}.
As \citet{Judge+al:2004} were mainly interested in the irradiance in the
various lines, and not in the actual shape of the full-Sun line profile,
this procedure is perfectly adequate.

In case one is interested in details of the line profile, e.g.\ the
superposition of a narrow and a broad component, one can no longer rely on
such a method as proposed by \citet{Judge+al:2004}.
Instead in the present paper a construction of a full-Sun spectrum based
on a full-disk raster scan of the SUMER instrument will be utilized.
Due to instrumental effects a simple summation of the individual spectra
acquired during the full-disk rasters is not possible, and a more
sophisticated approach will be used.

The outline of the paper is as follows.
In \sect{S:obs} the SUMER full-disk scans and instrumental effects are
described before the construction of the full-Sun spectrum is outlined.
The resulting full-Sun spectrum is characterized in \sect{S:character},
and center-to-limb effects and the influence of spatial structures are
discussed, before the consequences of these results are laid out in
\sect{S:interpretation}.
Finally the full-Sun spectrum is compared to {\aCenA} and other solar-like
stars in \sect{S:stars} as well as to luminous cool stars in
\sect{S:wind}, paying special attention to the consequences  for the
interpretation of these stellar profiles.
\sect{S:conclusions} concludes the paper with a discussion of the main
results.

\section{From a full-disk scan to a full-Sun spectrum} \label{S:obs}

\subsection{The SUMER C\,{\small{IV}} full-disk scan}  \label{S:sumer}

The SUMER VUV spectrograph acquired a number of full-disk scans in various
VUV emission lines, which have been analyzed in detail to investigate the
radiance of the Sun, e.g.\ with respect to the center-to-limb variation or
the spatial distribution of line intensities \citep{Wilhelm+al:1998}.
Out of these full-disk rasters only three were recorded containing the
full spectral information, i.e.\ the line profile; for the others just
moments of the profile have been transmitted to the ground, which is useful
to investigate the line radiance, but, of course, insufficient to study
details of the spectral profile.

The spectral properties of the three scans with full spectral information
in \ion{He}{1} (584\,\AA), \ion{Ne}{8} (770\,\AA) and \ion{C}{4}
(1548\,\AA) have been investigated in detail by
\cite{Peter:1999full,Peter:1999he}.
Using these data \cite{Peter:1999full} could show that the transition
region line shifts really show a center-to-limb variation in accordance
with a predominantly radial flow, i.e.\ the shift is proportional to the
cosine of the heliocentric angle, and that (low) coronal lines show a
net blueshift at disk center, while the transition region is redshifted.

The data reduction procedure for the SUMER data is described in detail in
\cite{Peter:1999full}; peculiarities of the data processing, e.g.\
correcting the cushion distortion of the image on the SUMER
detector are discussed thoroughly by \cite{Peter+Judge:1999}.
In short, one has to correct the data for flat-field, dead time, gain, and
geometric distortion as well as for the temporal drift of the image on the
detector, which will be of importance in the next subsection
(\sect{S:drift}).

In the present study the SUMER full-disk data as reduced by
\cite{Peter:1999full} are used to construct a full-Sun spectrum,
concentrating on the full-disk raster in \ion{C}{4}  (1548\,\AA).
Under ionization equilibrium conditions this line is formed
at ${\approx}10^5$\,K in the middle transition region.
The raster was performed on 4.\ and 5.\ Feb.\ 1996 over a total time of 31
hours.
The exposure time of each individual spectrum acquired through the
$1{\arcsec}\!{\times}300\,{\arcsec}$ slit was 15\,s, and a $\approx$2.1\,{\AA} (50
wavelength pixels) wide part of the spectrum centered around the \ion{C}{4}
line was transmitted to the ground.
The line was placed on reference pixel 403 on the KBr coated part of the
detector.
The scan step (perpendicular to the slit) to aquire the raster was
${\approx}$3{\arcsec}, so the resulting maps are (${\approx}3{\times}$)
under-sampled.
However, as all parts of the solar disk are equally represented in the
raster, this does not pose a problem when constructing a full-Sun
spectrum.
At the peak of the \ion{C}{4} line on average (median value) only 13
counts per pixel were registered.
Therefore the data have been binned along the slit by a factor of 3.

After the binning a single Gaussian fit was
applied to each spectral profile using a Genetic Algorithm based
optimization method \citep[{\sc{Pikaia}},][]{Charbonneau:1995}, which
proved to be a very robust and reliable tool for the problem at hand.

As there is no on-board wavelength calibration lamp with SUMER, the line
positions are calibrated so that the line shifts at and directly above the
limb vanish on average (when having corrected for solar rotation).
Above the limb the line of sight completely transverses the optically thin
transition region, and statistically one should see as much plasma flowing
towards the observer as is flowing away.
Thus there should be no net line shift.
This is discussed in detail by \cite{Peter:1999full} and
\cite{Peter+Judge:1999}.

\begin{figure}
\centerline{\includegraphics{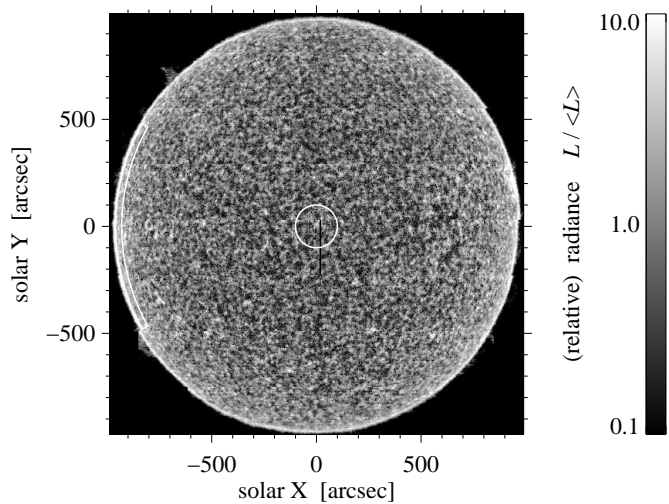}}
\caption{%
Line radiance map of the solar disk in \ion{C}{4} (1548\,\AA) relative to
the median value.
The solid lines near disk center and the east limb outline regions for
which average spectra will be shown in \fig{F:spectra}b,c.
North is top, east is left.
\label{F:intensity}}
\end{figure}

\begin{figure*}[!t]
\sidecaption
\includegraphics{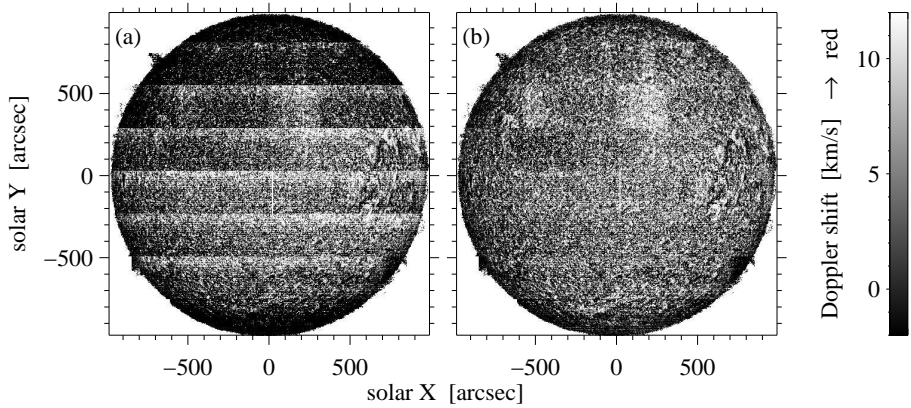}
\caption{%
Doppler shift map of the solar disk in \ion{C}{4} (1548\,\AA).
The left panel (a) shows the Doppler shifts before the corrections
described in \sect{S:sumer}, the right panel (b) shows the processed map.
In the final map (b) one can see the center-to-limb variation of
the Doppler shift and the signal of the solar rotation (${\approx}$1.8\,km/s
at the equator).
\label{F:shift}}
\end{figure*}

In \fig{F:intensity} the raster of the full Sun is plotted in line
radiance relative to the median radiance of the disk.
The Doppler shifts of the \ion{C}{4} line before and after all the
corrections are plotted in \fig{F:shift}a and b, respectively.
Please note the strong variation of line shift in the uncorrected Doppler
map (\fig{F:shift}a) while acquiring the data starting from the
south-east (bottom-left), first moving westward and finally finishing in
the north-east (top-left).
One can still identify the eight sweeps of the spectrograph
rastering the full disk back and forth in \fig{F:intensity} and
\fig{F:shift}, especially well in the uncorrected Doppler shifts
(\fig{F:intensity}a).

\subsection{SUMER long-term wavelength drifts}	\label{S:drift}

In principle one could construct the full-disk integrated spectral
profile of the \ion{C}{4} line by simply adding up all the individual
spectra acquired across the whole disk.
However, this does not work, as the SUMER instrument shows a systematic
temporal drift of the image on the detector in the wavelength direction,
which is related to the temperature stabilization of the instrument, kept
at a defined mean temperature during operation as has been shown in detail
by \cite{Rybak+al:1999}.
The typical time scale of this drift is about two hours with a
peak-to-valley variation of the line position on the detector of
about 0.7 spectral pixels, corresponding to 0.03\,{\AA} or 
about 6\,km/s Doppler shift for the \ion{C}{4} line at 1548\,{\AA}
\citep[see Fig.\,1 of][]{Rybak+al:1999}.
%


For longer observations with heavy usage of the (pointing) mechanisms, like
during the 31 hours full-disk raster under investigation, stronger drifts
can occur, in the present case summing up to ${\approx}0.8$ or more spectral
pixels, corresponding to Doppler shifts of $\pm$7\,km/s
\cite[cf.][]{Peter:1999full}.
This is nicely illustrated by the uncorrected Doppler map shown in
\fig{F:shift}a.
The long-term drift can be corrected for, based on the argument that the
variation of the Doppler shift should be systemantic on the Sun, i.e. on
average it should show a smooth variation from one limb through disk center
to the other limb.
Especially the east-west varaition should be the same as the north-south
variation.
These corrections for the full-disk Doppler shifts as shown in
\fig{F:shift}a \& b have been performed by \cite{Peter:1999full}.

In conclusion, the amplitude for the systematic {\em instrumental} drifts
of the line shifts, summing up to some 14\,km/s, is not negligible%
\footnote{Note that under normal circumstances for most of the
observations performed with SUMER, this instrumental drift of the line
position plays only a minor r\^ole or is completely negligible.
It is only the very long duration combined with the heavy use of the
mechanisms of the instrument for the full-disk scan which leads to this
strong effect.}
when compared to the width of a typical transition region line of some 20
to 30\,km/s (Gaussian width, i.e.\ half width at $1/e$ of maximum).
Therefore we cannot simply add up all the spectra acquired during the
full-disk scan to get a full-Sun spectrum, as the resulting spectrum would
be heavily influenced by the instrumental effects.

One possible solution would be to shift each individual spectrum
in wavelength by the same amount as the Doppler shifts are corrected.
However, the spectra, especially in the less bright inter-network
areas, are still relatively noisy, even after the binning, which is
because of the short exposure time and the corresponding low counts
(cf.\ \sect{S:sumer}).
In order to shift the line profile {\em reliably} by a fraction of a
pixel, one would have to be able to perform a reliable interpolation of
the spectral profile.
As this is prevented by the low signal-to-noise ratio, this strategy seems
problematic for the present data set.

\subsection{Constructing the full-Sun spectrum}	\label{S:constr}

Because of the problems outlined above, for the present study an
alternative procedure based on the Gaussian fits to the line profile will
be used.
While the original spectra have a too low signal-to-noise ratio in order
to shift them by a fraction of a pixel in wavelength through interpolation,
the data quality is sufficient for a reliable Gaussian fit to the line
profile using an appropriate tool.
Being a global optimization method, the Genetic Algorithm implemented by
\cite{Charbonneau:1995} is ideal for this task, and it has proven to give
reliable results on a large number of problematic optimization tasks.
Therefore we use the line radiances (cf.\ \fig{F:intensity}), corrected
Doppler shifts (cf.\ \fig{F:shift}), line widths and continuum levels as
following from the (single) Gaussian fits to calculate the line profiles
at each respective location on the solar disk.
As mentioned in the original work, the single Gaussian fits give an
excellent description of the original line profiles
\citep{Peter:1999full}.

\begin{figure*}
\sidecaption
\includegraphics{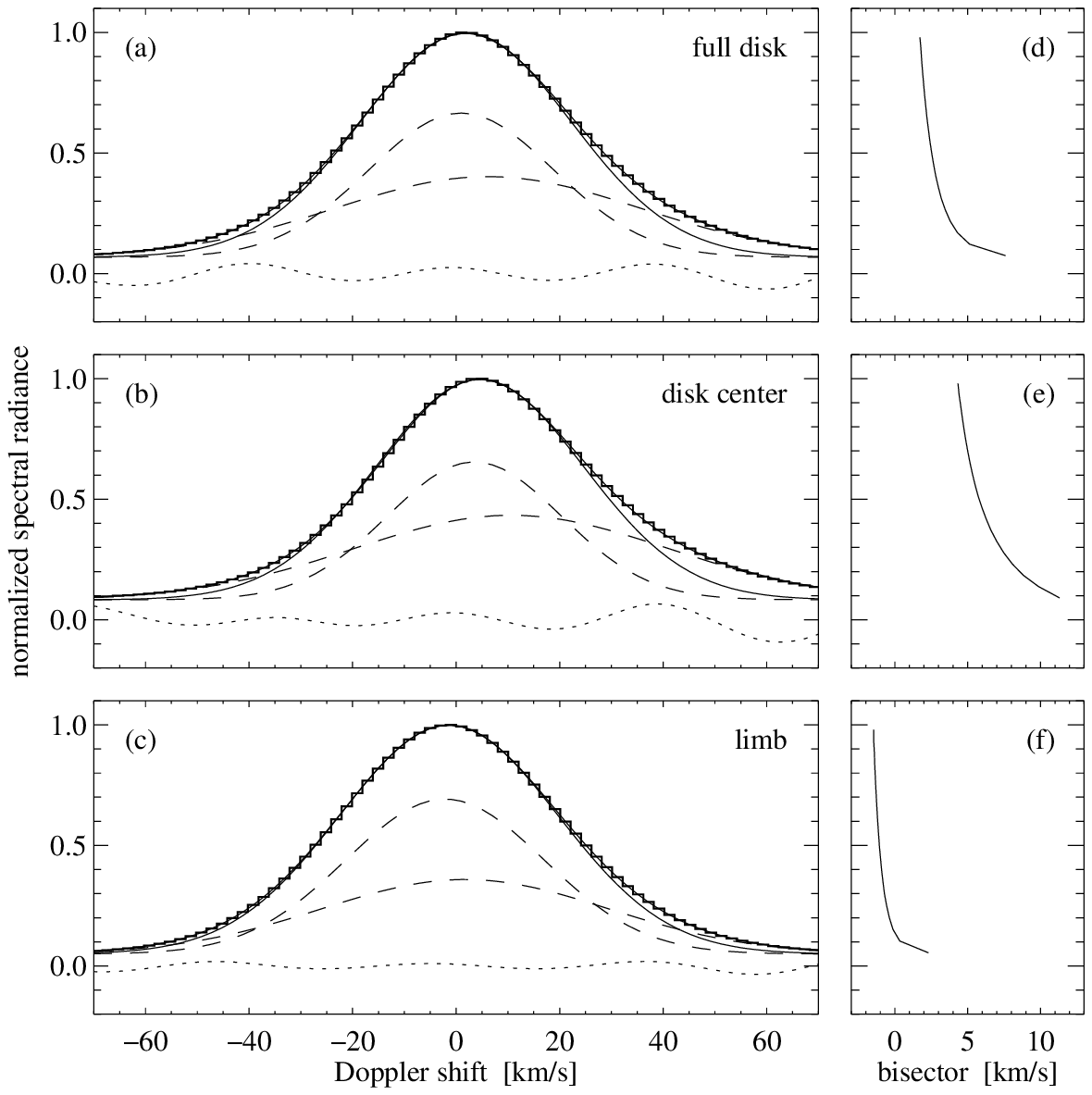}
\caption{%
~First VUV full-Sun line profile in \ion{C}{4} (1548\,\AA) with high
spectral resolution (panel a; solid line in histogram mode).
\newline
The dashed lines show the two components of a double Gaussian fit to
the line profile; the double Gaussian fit itself is practically identical
to the line profile.
The dotted line shows the relative difference between the double Gaussian
fit and the full disk spectrum (multiplied by 5); the relative difference
stays well below 1\,\%.
For comparison the solid line shows a single Gaussian fit.
\newline
Panels (b) and (c) show the same, but for the disk center and limb regions
as outlined in \fig{F:intensity}.
\newline
Panels (d) to (e) show the bisectors of the line profiles shown in panels
(a) to (c), respectively.
\label{F:spectra}}
\end{figure*}

These spectra can now be easily summed up to give a full-Sun spectrum,
or to evaluate spectra of smaller areas on the Sun for comparison.
In the radiance map (\fig{F:intensity}) we outlined one such region at
disk center (with a diameter of ${\approx}$1/10 of the disk diameter) and
another  region from 95\,\% of the disk radius (930${\arcsec}$) to the limb
(960${\arcsec}$) within $\pm30^\circ$ from the equator at the east limb.

The potential problem of this procedure is that only a single Gaussian fit
is performed to the line profiles, while it has been known for long that the
transition region lines, even when observed with the highest possible
spatial resolution (1{\arcsec} to 2{\arcsec}), partly show enhanced emission
in the wings of the lines \citep{Kjeldseth-Moe+Nicolas:1977}, which is well
described by a double Gaussian fit \citep{Peter:2000:sec:err}.
With the present data the signal-to-noise ratio due to the low count rates
prevents a double Gaussian fit.
However, it will be shown in \sect{S:high_res} that these broad components
describing the wings of the high spatial resolution spectra contribute
only little to the total emission, and therefore might be neglected for
the present investigation.

The full-Sun spectrum and, for comparison, the disk-center and limb
spectra are displayed in \fig{F:spectra}a--c as thick lines (in histogram
mode).
\fig{F:spectra}a shows the first VUV full-Sun line profile with a spectral
resolution allowing to investigate details of the line profile.
In contrast to earlier work this spectrum is not only based on spectra
taken on a small fraction of the visible solar hemisphere, but is obtained
through an integration over the full solar disk.

\section{Characteristics of the solar full-disk spectrum} \label{S:character}

To characterize the full-Sun spectrum and to compare it with stellar
spectra, a single and a double Gaussian fit is applied to the spectrum.
Here we use again the Genetic Algorithm based fit \citep{Charbonneau:1995}
together with a fast local method (Levenberg-Marquardt; IDL-code
{\tt mpfit} implemented by C.B. Markwardt).
The fits are shown in \fig{F:spectra}a and the fit parameters are listed in
\tab{T:params}.

\begin{table}[t]
\newcommand{\NC}{{{NC}}}
\newcommand{\BC}{{{BC}}}
\newcommand{\SG}{{\it{SG}}}
\caption{Parameters for \ion{C}{4} (1548\,\AA) of the full-Sun spectrum as
well as the disk-center and limb spectra as plotted in \fig{F:spectra}.
For the single Gaussian fits (\SG) as well as for the narrow and broad
components (\NC,\BC) of the double Gaussian fits the columns list the
radiance of the respective component compared to the total line radiance
$I/I_{\rm{line}}$, the Doppler shifts $v_{\rm{D}}$ (positive $=$ red), and
the line widths, which are given as Gaussian widths $w_{1/e}$, full width
at half maximum {\FWHM} and the resulting non-thermal broadening $\xi$.
For comparision, the table also lists the parameters for solar line
profiles from bright network elements observed in
$1{\arcsec}\!{\times}1{\arcsec}$ large pixels near disk center
taken from \cite{Peter:2001:sec} and the line fit parameters for the
$\alpha$Cen\,A spectrum following \cite{Pagano+al:2004}.
\label{T:params}}
\begin{center}
\begin{tabular}{@{}lccrcccccccc@{}}
\hline
\hline
\\[-1.9ex]
\multicolumn{2}{l}{\ion{C}{4} (1548\,\AA)}
&
& $v_{\rm{D}}$~~~
& \multicolumn{3}{c}{line width  [km]}
\\
\cline{5-7}
&
& \raisebox{1.4ex}[-5ex]{$\displaystyle\frac{I}{I_{\rm{line}}}$}
& [km/s]
& $w_{1/e}$
& \FWHM
& $\xi$
\\[0.5ex]
\hline
\\[-1.5ex]
full Sun   & \SG & \it 0.92 & \it    1.9 ~ & \it 28.7 & \it 47.7 & \it 26.1 \\
           & \NC &     0.52 &        1.0 ~ &     25.3 &     42.1 &     22.4 \\
           & \BC &     0.48 &        6.1 ~ &     41.8 &     69.6 &     40.2
\\[0.9ex]
disk center& \SG & \it 0.90 & \it    4.6 ~ & \it 28.3 & \it 47.1 & \it 25.8 \\
           & \NC &     0.47 &        3.5 ~ &     23.8 &     39.6 & \it 20.7 \\
           & \BC &     0.53 &       10.6 ~ &     43.0 &     71.5 &     41.3 
\\[0.9ex]
limb       & \SG & \it 0.96 & \it $-$1.4 ~ & \it 29.7 & \it 49.5 & \it 27.3 \\
           & \NC &     0.59 &     $-$2.1 ~ &     27.1 &     45.1 &     24.4 \\
           & \BC &     0.41 &        1.7 ~ &     39.3 &     65.4 &     37.5
\\[0.5ex]
\hline
\\[-1.5ex]
$1{\arcsec}\!{\times}1{\arcsec}$ bright
           & \SG & \it 0.79 & \it    8.0 ~ & \it 30.7 & \it 51.1 & \it 28.4 \\
network    & \NC &     0.72 &        8.3 ~ &     25.2 &     41.9 &     22.3 \\
spectrum   & \BC &     0.28 &        7.2 ~ &     53.4 &     88.9 &     52.1
\\[0.5ex]
\hline
\\[-1.5ex]
$\alpha$Cen\,A
           & \NC &     0.53 &        8.8 ~ &     25.9 &     43.2 &     23.2 \\
           & \BC &     0.47 &        7.4 ~ &     47.3 &     78.8 &     45.8
\\[0.5ex]
\hline
\end{tabular}
\end{center}
\end{table}

The double Gaussian fit is an excellent match to the line profile, with
the relative difference to the full-Sun spectrum being well below 1\,\%
(the dotted line in \fig{F:spectra}a shows the relative difference
multiplied by 5).
The line profile shows a significant asymmetry, as the broad component of
the double Gaussian fit is shifted relative to the narrow component by
some 5\,km/s, which is a sizeable fraction of the width of the narrow
component.
The narrow and the broad component contribute roughly equal shares to the
total emission in the line.

\subsection{Differences between the full-disk spectrum
            and average disk-center and limb spectra}  \label{S:full_disk}

The full-Sun line profile and the one originating from a small region
around disk center (cf.\ \fig{F:intensity}) as plotted in \fig{F:spectra}a
and b look very similar.
The Gaussian fits for full-Sun and disk-center spectra are also very
much alike in terms of line radiance and line width, but not in line
shift (cf.\ \tab{T:params}).

The (east) limb spectrum (\fig{F:spectra}c) is much more symmetric
compared to the disk center one, as can also be seen from the bisectors of
the line profile in \fig{F:spectra}d--f.
Furthermore the spectrum at the limb shows a weaker and less shifted broad
component.
The line width there is somewhat larger than at disk center, which can be
understood in terms of an opacity effect, i.e.\ \ion{C}{4} is not
completely optically thin near the limb, where the path length of the
line of sight through the transition region becomes
longer \citep[cf.][Sect.\ 5.2.4]{Mariska:1992}.
However, as this effect is small, it affects the full-Sun line
profile only little.

The most noticeable difference between the disk-center and the limb spectra
is the line shift.
While at disk center \ion{C}{4} shows the well known redshift
\citep{Doschek+al:1976}, the line at the limb should show no shift
at all, as \ion{C}{4} is formed (almost) under optically thin
conditions.
However, because of the rotation of the Sun with some 1.8\,km/s at the
equator, the spectra at the east and west should show the corresponding
line shifts.
Indeed, the line profile is quite symmetric, relatively well
described by a single Gaussian fit (cf.\ \fig{F:spectra}c), and it is
blueshifted, as expected for the east limb (cf.\ \tab{T:params}).

The resulting line shift of the full-Sun spectrum should be somewhere
between zero and the disk center shift, but the actual value depends on the
relation of intensity to Doppler shifts and the center-to-limb variation
of both quantities.

Usually the shifts of transition region lines from the Sun are given as
average line shift at quiet Sun disk center.
This was the case in the first report of persistent transition region line
shifts by \cite{Doschek+al:1976} and is also true for the recent and still
up-to date SUMER quiet Sun investigations by \cite{Brekke+al:1997} and
\cite{Peter+Judge:1999} reporting redshifts of \ion{C}{4} from 4.5 to
6\,km/s as well as a study of a (moderately) active region by
\cite{Teriaca+al:1999:ar} giving a slightly higher redshift.
One should note that these authors derived their line shifts by fitting
the line profiles by a \emph{single} Gaussian.

In accordance with these results the \ion{C}{4} line (when fitted by a
single Gaussian) is redshifted by 4.6\,km/s at disk center (cf.\
\tab{T:params}).
In contrast, (the core of) the full-Sun spectrum shows a considerably
smaller redshift, only 1\,km/s or 2\,km/s, depending on whether one used a
double Gaussian fit or a single Gaussian fit (which is then pulled to the
red by the line asymmetry in the wing, which is accounted for by the double
Gaussian fit).

This has an important implication for stellar studies.
So far, usually  the disk-center quiet-Sun values have been used for
solar--stellar comparisons (cf.\ \sect{S:stars}).
However, as the stellar spectra are always disk-integrated, one has to
compare the stellar redshift not to the solar disk-center values, but to
values reduced by a factor of 2 to 4!
Using the persistent redshifts of the transition region lines as an
indicator of stellar activity, one vastly overestimates the activity of the
Sun when using the ``traditional'' disk center values.

\subsection{Differences between the full-disk spectrum
            and spectra at high spatial resolution}  \label{S:high_res}

Spectra of \ion{C}{4} on the solar disk with the highest spatial
resolution currently available (1{\arcsec} to 2{\arcsec}) have been
investigated by \cite{Peter:2000:sec:err,Peter:2001:sec} with respect to
the line asymmetries through double Gaussian fits.
These typically show line widths and shifts of the line core component
(the narrow component) comparable to the disk-center spectrum listed in
\tab{T:params}.
However, with respect to the broad components, there are three major
differences of the high spatial resolution spectra when compared to the
disk-center spectrum presented here%
\footnote{%
The 100{\arcsec} radius disk-center region used here (cf.\
\fig{F:intensity}) corresponds to 1\,\% of the solar disk.  The results do
not change much if one chooses a region at disk center with 200{\arcsec} or
even 300{\arcsec} radius, the latter one covering 10\,\% of the disk.
Sticking to the 100{\arcsec} radius region at disk center, however,  is
still large enough to be representative of the quiet Sun, but avoids any
center-to-limb effects.}
(cf.\ \tab{T:params}).
\\
(1) The broad components of the high spatial resolution spectra from bright
    network elements contribute only some 25 to 30\,\% to the total line
    radiance, while the broad component of the disk-center spectrum
    contributes 50\,\%, i.e.\ it is as strong as the narrow component.
\\
(2) In the high spatial resolution spectra the broad component is shifted
    slightly by some 1\,km/s to the {\em blue} with respect to the narrow
    component, while in the disk-center spectrum the relative shift of the
    broad component is about 7\,km/s to the {\em red}.
\\
(3) The broad components in the high spatial resolution spectra are much
    broader than in the disk-center spectrum (by ${\approx}$25\,\%).

Especially the first point is of interest in order to check the importance
of the broad components of the high spatial resolution spectra for the
calculation of the full-Sun spectrum.
According to \cite{Peter:2000:sec:err} the enhanced emission in the wings of
high spatial resolution spectra, accounted for by the broad components, is
mostly restricted to the bright patches of the network covering some 13\,\%
of the area.
Using this and the numbers given in Table 2 of
\cite{Peter:2000:sec:err}, the broad components of the high spatial
resolution spectra contribute only some 10\,\% to the total emission of the
line.

Thus the broad components of the high spatial resolution spectra could only
contribute a small (of order 20\,\%) portion to the broad  component of the
full-Sun spectrum, which accounts for 50\,\% of the total line emission.
In retrospect, this justifies to neglect the emission from the wings of
the \ion{C}{4} line when constructing the full-Sun spectrum
(cf.\ \sect{S:constr}).

\section{Interpretation of disk-integrated spectra} \label{S:interpretation}

The above discussion has major consequences for the interpretation of the
full-Sun spectrum and thus also for the interpretation of stellar VUV
spectra from transition region lines.

The tail components, or broad components, of the spectra at high spatial
resolution are certainly a signature of not-resolved small scale processes
on the Sun.
Explosive events, small reconnection events in the transition region, are
a good candidate to explain these broad components, as was already argued
by \citet{Dere+Mason:1993}.
The larger of these events, when spatially resolved, cause satellites in
the blue and/or red wing of the line, shifted by up to 100\,km/s or more
due to an outflow from the reconnection site \citep{Innes+al:1997}.
Smaller explosive events, not spatially resolved, could then lead to the
enhancement in the line wings, even though \cite{Peter+Brkovic:2003}
presented some counter-arguments.
An alternative, still speculative scenario was proposed by
\citet{Peter:2001:sec}, in which the broad components are interpreted as the
signature of of upward propagating magneto-acoustic waves in coronal
funnels, based on an investigation of VUV lines covering the whole
transition region and low corona.

Now the important question is, if one can use these interpretations for
the broad components of the high spatial resolution spectra of the Sun
also to understand the broad components of other stars.
Considering the discussion in the preceeding subsection, the answer might
be ``no''.

In the present analysis we have started with perfectly single Gaussian
line profiles derived from the actually observed spectra.
Therefore, the broad components of the full-Sun spectrum cannot be due to
broad components of the individual high spatial resolution spectra.

To investigate the nature of the broad component of the full-Sun spectrum
it is instructive to first look at the disk-center spectrum, which shows
very similar properties than the full-Sun spectrum.
Statistically, on the Sun the redshifts are strongest over the
magnetic chromospheric network, where also the transition region
intensities are the highest
\citep{Brynildsen+al:1996,Peter:1999full,Peter:2000:sec:err}.
This relation of line radiance and shift can well lead to a line
asymmetry with enhanced emission in the red wing, caused by the few
brightest areas showing very high redshifts.
This would also explain why the enhancement in the red wing is much stronger
than in the blue wing, viz.\ the redshift of the broad component as
compared to the narrow one.
Please note that the Gaussian widths of the broad components of the full-Sun
and disk-center spectra (cf.\ \tab{T:params}) are just below the
(adiabatic) sound speed of ${\approx}$50\,km/s at the formation temperature
of \ion{C}{4} of $10^5$\,K and thus the motions leading to the broad
components are subsonic.

Finally, the much reduced shift of the narrow and core components of the
full-Sun spectrum as compared to the one at disk center is simply a
consequence of the center-to-limb variation and the (statistically)
vanishing line shifts at the limb (cf.\ \sect{S:full_disk}).

The conclusion from this discussion is that the enhanced emission in the
wings of the full-Sun spectrum is {\em not} due to un-resolved
small-scale processes like explosive events or wave propagation.
It is more likely that it is primarily due to the large scale structures
of the super-granular convection, viz.\ the chromospheric network, with a
typical scale of some 3 to 5\,\% of the solar radius.
This is based on the systematic differences in the flow-structure of the
network and inter-network areas and especially the relation of the line
shifts to radiances \citep[e.g.][]{Brynildsen+al:1996,Peter:2000:sec:err}.
As the width of the broad component is the same in the disk-center and the
full-Sun spectrum, the broad component of the full-Sun spectrum still
carries information on the structure and size of the chromospheric
network.

From previous work it is well established that the width of the narrow
components of the disk-integrated spectra contain valuable information on
the (coronal) heating process, as discussed e.g.\ in \cite{Wood+al:1997},
who compare a large number of stars width different activity levels.
Using this information one might derive some estimate on the average net
line shift.
Through this the Doppler shift of the narrow component of the
disk-integrated spectrum then contains information on the center-to-limb
variation of the transition region emission.

From this one might distill the following rough trends concerning the
disk-integrated spectrum:
\\
(1) The broad component contains information on the structure of and flows
    in the chromospheric network.
\\
(2) The width of the narrow component contains information on the stellar
    activity and the heating process.
\\
(3) The Doppler shift of the narrow component contains information on the
    center-to-limb-variation.

\section{Comparison to {\boldmath$\alpha$\unboldmath}Cen A\newline
         and other solar-like stars}  \label{S:stars}

As the Sun and {\aCenA} are quite similar in spectral type (both G2\,V),
and as {\aCenA} has a very high apparent luminosity, {\aCenA} seems to be
a very good choice for a solar analogue.
For this reason and especially because of the lack of a full-Sun
VUV spectrum with decent spectral resolution, \cite{Pagano+al:2004}
proposed to use the {\aCenA} spectrum as a reference for solar-stellar
connections.
The line fit parameters for \ion{C}{4} (1548\,\AA) from {\aCenA} as given
by \cite{Pagano+al:2004} are repeated in \tab{T:params}.
However, there are also indications for differences between the Sun and
{\aCenA}, e.g.\ a sightly higher mass and a ${\approx}$25\,\% lower surface
gravity \citep{Morel+al:2000}.
The most noticeable difference in terms of magnetic activity is the lack
of a clear indication for a UV cycle with {\aCenA}, at least on a time
scale comparable to the Sun \citep{Ayres+al:1995}.
Therefore it is of vital importance to compare the full-Sun spectrum of
\ion{C}{4} (1548\,\AA) to {\aCenA}.

Both the full-Sun and the {\aCenA} line profile are well described by a double
Gaussian fit, and the relative contributions of the two components match
very well (cf.\ \tab{T:params}).
Likewise the widths of the narrow and broad components differ only by
${\approx}$2\,\% and ${\approx}$12\,\%, respectively.
In that respect {\aCenA} indeed is a very good representative of the Sun.

There is, however, a very pointed difference in the line shifts for the
full-Sun and {\aCenA} spectra, manifesting itself by two findings:
(1) The {\aCenA} narrow component shows a more than four times higher
    redshift than the narrow component of the full-Sun spectrum.
    Previous studies could not find this disagreement because they compared
    the {\aCenA} spectrum to solar spectra at high spatial
    resolution, rather than to a disk-integrated spectrum as derived in
    this paper.
%
%
(2) While both the full-Sun and the {\aCenA} broad component show a
    noticeable redshift, in the case of the full-Sun spectrum the broad
    component shows a \emph{redshift} relative to the narrow component,
    while for {\aCenA} there is a small relative \emph{blueshift}.

According to the discussion in \sect{S:interpretation} this difference in
line shifts of the narrow and core components might hint at differences
between the Sun and {\aCenA} with respect to the structure of the
chromospheric magnetic network and the center-to-limb variation
(items 1 and 3 at the end of \sect{S:interpretation}).
Considering the difference in surface gravity between the Sun and {\aCenA}
we might well expect a  difference in the super-granular convection pattern
--- changing the network structure --- and in pressure scale height ---
changing the emission scale height of the transition region and thus the
center-to-limb variation of \ion{C}{4}.


\section{Investigating the line asymmetry:\newline
         Diagnostics for stellar winds?}		\label{S:wind}

In their detailed investigation of VUV line profiles from luminous cool
stars \cite{Dupree+al:2005} propose that the winds of these stars lead
to a deficit in the blue part of the line profile through an opacity effect.
They investigated \ion{C}{3} (977\,\AA) and \ion{O}{6} (1032\,\AA), formed
in ionization equilibrium conditions at ${\approx}80\,000$\,K and
${\approx}240\,000$\,K, respectively.
The \ion{C}{3} profiles from the various stars are reasonably well fitted
with a single Gaussian (within the noise of the data; except for the
inter-stellar absorption feature near line center).
For the \ion{O}{6} line a double Gaussian fit with a narrow and
an approximately equally strong broad component is necessary.

If the proposal of \cite{Dupree+al:2005} is correct, together with
future modeling of the respective stellar wind this would provide stellar
physics with a new powerful tool to investigate the mass loss rate of
luminous cool stars.
Therefore it is of importance to investigate their assumption that the
line asymmetries are indeed a signature of the opacity of the wind
outflow.

Assuming that the respective VUV emission line is formed close to the
star, where the plasma is still almost at rest, and assuming a wind which
is \emph{not} optically thin, one would expect a deformation of the line
profile due to absorption in the blue part of the line profile.
Based on this idea \cite{Dupree+al:2005} do a single Gaussian fit to only
the red side of the line profile, which will not (or only little) be
affected by absorption in the wind.
The excess in the blue side of the fitted Gaussian compared to the observed
spectrum is then a measure for the line emission lost from the
line profile due to the opacity effect of the wind \cite[cf.\ Figure 9 and
12 in][showing the \ion{C}{3} and \ion{O}{6} spectra along with the
fits]{Dupree+al:2005}.

\begin{figure}
\includegraphics{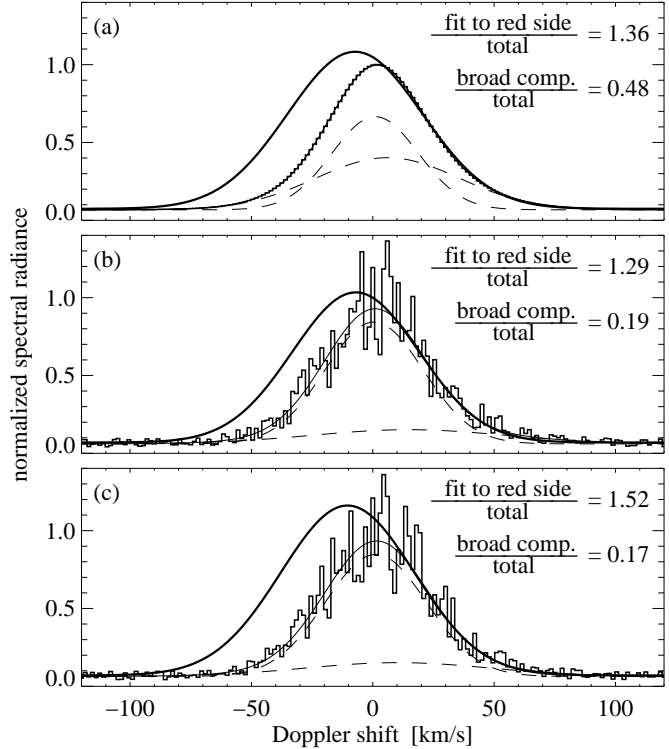}
\caption{%
Single Gaussian fits to the red side of the line profiles only.
Shown are the the full-Sun spectrum (panel a) and two cases of the full-Sun
spectrum with 40\,\% noise added (panels b, c).
The spectra are displayed in histogram mode, the fits as thick solid lines.
For comparison, double Gaussian fits to the whole profiles are shown as thin
solid lines with the dashed lines indicating the two Gaussian components.
The numbers with the plots show the radiance of the single Gaussian fit
to the total radiance of the spectrum as well as the contribution of the
broad component of the double Gaussian fits.
See \sect{S:wind}
\label{F:noise}}
\end{figure}

To investigate their proposal, the same analysis procedure is applied to
the full-Sun spectrum shown in \fig{F:spectra}a.
The spectrum is plotted again in a histogram style in \fig{F:noise}a, now
along with a single Gaussian fit using only the red side of the line profile
(positive Doppler shifts) to perform the fit.
The resulting Gaussian is over-plotted as a thick solid line.
It is evident that the solar spectrum shows the same effect as the
spectra of the luminous cool stars reported by \cite{Dupree+al:2005}.

To have a more realistic comparison, we binned the full-Sun spectra to
roughly match the spectral resolution of the FUSE spectra used by
\cite{Dupree+al:2005}, and added some 40\,\% random noise.
\fig{F:noise}b and c show two cases with noise on top of the same
full-Sun spectrum in histogram mode. The thin solid lines show double
Gaussian fits to the noisy spectra, with the two components as dashed
lines.
The single Gaussian fit to only the red part of the profile is over-plotted
as a thick solid line.
Typically this single Gaussian has an integrated radiance which is some
30\,\% to 50\,\% higher than for the actual full-Sun profile (cf.\ \fig{F:noise}).

One result of this simple experiment is that the noisy spectra
(\fig{F:noise}b,c) show a much weaker broad component than the original
full-Sun spectrum (\fig{F:noise}a).
While in the latter case the broad and narrow components are equally strong
(the broad component contributes 47\,\% to the total emission), in the case
of the noisy spectra the broad component contributes less than 20\,\% to the
total emission.
This result is natural, as for a noisy spectrum the narrow component can be
pushed to larger line widths by the optimization algorithm and thus there
is less need for a broad component when performing the line fit.
This effect might be part of the reason why \cite{Dupree+al:2005} found
\emph{single} Gaussians to be sufficient to fit the \ion{C}{3} line they
show in their Fig.\ 9.

The major result of this experiment is that the noisy line profiles,
\emph{and} their fits to the red side only of the line profile, look very
similar to the cool luminous stars \citep[Fig.\ 9 \& 12
of][]{Dupree+al:2005}.
This is especially true for $\alpha$\,Car (F0\,II),
$\beta$\,Cet (K0\,III) and $\beta$\,Gem (K0\,III).

In the case of the Sun with its hot, low density wind, there is no opacity
effect the wind is imposing on the VUV spectra originating from the low
corona and transition region.
As outlined in the previous sections, the asymmetries in the full-Sun
spectra are due to the distribution of structures, especially the
chromospheric magnetic network.
Because of the close proximity of the full-Sun spectra with added noise and
the cool star spectra shown by \cite{Dupree+al:2005} with respect to the
single Gaussian fit to the red part of the line, one might wonder about the
validity of the assumption to start with, i.e.\ the opacity effect of the
wind.
Of course, their working hypothesis is very attractive, as one can expect a
strong wind from cool luminous stars.
However, this interpretation is not unique, as also noted by the authors.

For the Sun the asymmetry of the line is due to the cell-network structure,
with the strongest Doppler shifts occuring in the bright areas.
The net shift of transition region lines is, e.g., due to flows or waves
along loops, as has been suggested by various 1D loop models \citep[e.g.\
see references in][]{Peter+Judge:1999} and confirmed by a recent
investigation of spectra synthesized from a 3D MHD coronal model
\citep{Peter+al:2004}.

This situation could be similar to the cool, more luminous stars, where
one certainly would expect some larger scale convection patterns
\citep{Schwarzschild:1975,Freytag+al:2002} leading to
concentrations of the magnetic fields.
Like the magnetic network on the Sun, these concentrations will give rise
to a connection between line shift and radiance, then resulting in the
asymmetry of the integrated profile.
Furthermore also those cool luminous stars can be expected to have closed
magnetic structures in their low coronae and thus it is very probable that
the transition region and low corona shows the persistent line shifts known
very well for the Sun and other stars \citep{Wood+al:1997,Pagano+al:2004}.
The net line shifts (of the narrow components) given by
\cite{Dupree+al:2005} for the cool luminous stars range from zero to
20\,km/s to the red, which does not seem unrealistic considering values
found on the Sun, where the distribution of redshifts in the quiet Sun
transition region reaches up to at least 25\,km/s \citep[Fig.\ 9
  of][]{Peter:1999full}.

The conclusion from this is that there are at least two possible scenarios
to understand the nature of the asymmetries of VUV lines from cool
luminous stars.
The hypothesis of \cite{Dupree+al:2005} that opacity effects in the stellar
wind lead to an absorption in the blue part of the line, and the
suggestion put forward in this paper, that surface structures are
responsible for the effect.
Future modeling efforts on the distribution of structures in the
atmospheres of cool luminous stars as well as modeling of the radiative
transfer of the VUV lines through the stellar wind will be needed to
distinguish between these two suggestions, or to favor even another idea,
not yet thought of.

\section{Discussion and Conclusions}		\label{S:conclusions}

This paper presents the first VUV spectrum of the Sun seen as a star of a
transition region line with sufficient spectral resolution to allow a
detailed comparison with stellar observations.
To construct the full-Sun spectrum, a SUMER full-disk raster scan was used.
Before summing up all the spectra, instrumental corrections have been
applied, especially for the long-term wavelength drifts.
Through this procedure we find an asymmetric full-Sun spectrum with
enhanced emission in the wings of the line, which is very well fitted by a
double Gaussian with a narrow and a broad component.

From the comparison of the full-Sun spectrum with average and
high spatial resolution spectra from various locations on the solar disk we
find that the broad component of the full-Sun spectrum is \emph{not} a
signature of the enhanced emission in the wings of high spatial resolution
spectra, which are due to small scale reconnection flows or waves.
In contrast, full-Sun broad components are indicative of the structure of
the magnetic (chromospheric) network through the relation of intensity
and Doppler shift of transition region emission lines.
Therefore, unlike as proposed in previous stellar studies by e.g.\
\cite{Wood+al:1997} or \cite{Pagano+al:2004}, the broad components of
disk-integrated stellar spectra do \emph{not} contain much information
on the coronal heating mechanism, but rather provide valuable input on the
structure of the magnetic network.
This opens a new range opportunities for stellar studies.

The net redshifts of transition region lines of solar-like stars have been
used as an indication of stellar activity, because we know from the Sun that
active regions show much stronger redshifts than quiet areas
\citep{Achour+al:1995,Teriaca+al:1999:ar}.
However, the redshift of the full-Sun profile of \ion{C}{4} was found to be
lower by a factor of 4 to 6 when compared to data at disk center.
Thus, at least for the Sun, the center-to-limb variation greatly reduces
the line shift of the full-Sun spectrum.
This is because the intensity increases towards the limb, while the
line shift decreases \citep{Peter:1999full}.

When comparing the Sun to {\aCenA}, which shows a net redshift of almost
10\,km/s while the full-Sun spectrum is only shifted by 1--2\,km/s, there
are basically two interpretations.
Either {\aCenA} has a comparable disk-center line shift but no
center-to-limb variation, or {\aCenA} has a considerably higher disk-center
line shift than the Sun.
As it is hard to imagine {\aCenA} having no limb brightening in the
optically thin transition region lines, the second interpretation is more
plausible.
As the broad components of {\aCenA} and the full-Sun are comparable,
according to the above discussion the two stars should have similar network
structures.
Thus we can expect {\aCenA} to have much stronger redshifts above the
magnetic concentrations of the chromospheric network, and therefore
probably more magnetic flux concentrated in the magnetic network.

The comparison of the full-Sun spectrum to those of luminous cool stars
reveals that the interpretation of the latter by \cite{Dupree+al:2005}
as being due to an opacity effect in the wind is not unique.
While in the case of the Sun we can be sure that there is (almost) no
opacity effect in the wind, the full-Sun spectrum shows the same properties
as the luminous cool stars when the same techniques as used by
\cite{Dupree+al:2005} are applied.

These comparisons of the full-Sun VUV spectrum of \ion{C}{4} (1548\,\AA)
show that full-Sun spectra in other lines are needed.
The technique applied in this paper will work only for \ion{Ne}{8}
(770\,\AA) and \ion{He}{1} (584\,\AA), as SUMER acquired full-disk rasters
with full spectral resolution in these lines only.
To get profiles for a larger number of VUV emission lines, one would have
to apply a different technique.
The straylight analysis mentioned in the introduction, performed by
\cite{lemaire+al:2004} and so far only applied to continuum data, is a good
candidate.

Probably the most important result of this paper is to demonstrate the
necessity of complex modeling of the stellar structures, as only this can
provide us with the tools to investigate the similarities and differences
of solar and stellar VUV spectra.
This incorporates new simulations of solar and stellar coronae including the
synthesis of VUV spectra as done by \cite{Peter+al:2004}, as well as new
models of winds e.g.\ from cool luminous giants including radiative
transfer through the wind in order to distinguish between surface structure
effects and absorption in a dense wind.


\begin{acknowledgements}
Sincere thanks are due to Klaus Wilhelm and Reiner Hammer for carefully
reading the manuscript.
\end{acknowledgements}


\def\apj       {ApJ}%
\def\apjl      {ApJ}%
\def\apjs      {ApJS}%
\def\apss      {Astrophys. Space Sci.}%
\def\jgr       {J. Geophys. Res.}%
\def\nat       {Nature}%
\def\solphys   {Solar Phys.}%
\def\philtrans {Phil.\ Trans.\ Roy.\ Soc.\ Lond.}%
\def\aap       {A\&A}%
\def\aaps      {A\&A Suppl.}%
\def\mnras     {Mon.\ Not.\ Roy.\ Astron.\ Soc.}
\def\araa      {Ann.\ Rev.\ Astron.\ Astrophys.}
\def\naturwiss {Naturwiss.}
\def\za        {Zeitschrift f.\ Astrophys.}
\def\an        {Astronomische Nachrichten}
\def\pasp      {PASP}


\end{document}